\documentclass[fleqn,usenatbib]{mnras}

\usepackage{graphicx}	
\usepackage{amsmath}	
\usepackage{amssymb}	
\usepackage{multicol}        
\usepackage{bm}		
\usepackage{pdflscape}	
\usepackage{lipsum}
\usepackage{subcaption}
\usepackage{float}
\usepackage{longtable}
\usepackage[utf8]{inputenc}
\usepackage{comment}
\usepackage{lineno}

\usepackage{eso-pic}

\AddToShipoutPictureBG*{%
  \AtPageUpperLeft{%
    \hspace{0.75\paperwidth}%
    \raisebox{-3.5\baselineskip}{%
      \makebox[0pt][l]{\textnormal{DES-2022-0738}}
}}}%

\AddToShipoutPictureBG*{%
  \AtPageUpperLeft{%
    \hspace{0.75\paperwidth}%
    \raisebox{-4.5\baselineskip}{%
      \makebox[0pt][l]{\textnormal{FERMILAB-PUB-23-154-PPD}}
}}}%

\usepackage{newtxtext,newtxmath}

\usepackage[T1]{fontenc}

\DeclareRobustCommand{\VAN}[3]{#2}
\let\VANthebibliography\thebibliography
\def\thebibliography{\DeclareRobustCommand{\VAN}[3]{##3}\VANthebibliography}

\usepackage{graphicx}	
\usepackage{amsmath}	

\title[SIDM and Central Cluster Galaxy Offsets]{Examining the Self-Interaction of Dark Matter through Central Cluster Galaxy Offsets}

\author[DES Collaboration]{
\parbox{\textwidth}{
\Large
D.~Cross,$^{1,2,3}$\thanks{E-mail: cross@ice.csic.es}
G.~Thoron,$^{1,2}$\thanks{E-mail: gthoron@ucsc.edu}
T.~E.~Jeltema,$^{1,2}$\thanks{E-mail: tesla@ucsc.edu}
A.~Swart,$^{1,2,4}$
D.~L.~Hollowood,$^{1,2}$
S.~Adhikari,$^{5}$
S.~Bocquet,$^{6}$
O.~Eiger,$^{1,2,7}$
S.~Everett,$^{8}$
J.~Jobel,$^{1,2}$
D.~Laubner,$^{1,2}$
A.~McDaniel,$^{9}$
M.~Aguena,$^{10}$
O.~Alves,$^{11}$
F.~Andrade-Oliveira,$^{11}$
D.~Bacon,$^{12}$
E.~Bertin,$^{13,14}$
D.~Brooks,$^{15}$
D.~L.~Burke,$^{16,17}$
A.~Carnero~Rosell,$^{18,10,19}$
M.~Carrasco~Kind,$^{20,21}$
R.~Cawthon,$^{22}$
M.~Costanzi,$^{23,24,25}$
L.~N.~da Costa,$^{10}$
M.~E.~S.~Pereira,$^{26}$
T.~M.~Davis,$^{27}$
S.~Desai,$^{28}$
P.~Doel,$^{15}$
I.~Ferrero,$^{29}$
J.~Frieman,$^{30,31}$
J.~Garc\'ia-Bellido,$^{32}$
G.~Giannini,$^{33}$
D.~Gruen,$^{34}$
R.~A.~Gruendl,$^{20,21}$
S.~R.~Hinton,$^{27}$
K.~Honscheid,$^{35,36}$
D.~J.~James,$^{37}$
K.~Kuehn,$^{38,39}$
J.~L.~Marshall,$^{40}$
J. Mena-Fern{\'a}ndez,$^{41}$
F.~Menanteau,$^{20,21}$
R.~Miquel,$^{42,33}$
R.~L.~C.~Ogando,$^{43}$
A.~Pieres,$^{10,43}$
A.~A.~Plazas~Malag\'on,$^{44}$
M.~Raveri,$^{45}$
A.~K.~Romer,$^{46}$
E.~Sanchez,$^{41}$
I.~Sevilla-Noarbe,$^{41}$
M.~Smith,$^{47}$
M.~Soares-Santos,$^{11}$
F.~Sobreira,$^{48,10}$
E.~Suchyta,$^{49}$
M.~E.~C.~Swanson,$^{50}$
G.~Tarle,$^{11}$
C.~To,$^{35}$
N.~Weaverdyck,$^{11,51}$
J.~Weller,$^{52,53}$
and P.~Wiseman$^{47}$
\begin{center} (DES Collaboration) \end{center}
}}

\date{Accepted XXX. Received YYY; in original form ZZZ}

\pubyear{2022}

\begin{document}
\label{firstpage}
\pagerange{\pageref{firstpage}--\pageref{lastpage}}
\maketitle

\begin{abstract}
While collisionless cold dark matter models have been largely successful in explaining a wide range of observational data, some tensions still exist, and it remains possible that dark matter possesses a non-negligible level of self interactions. In this paper, we investigate a possible observable consequence of self-interacting dark matter: offsets between the central galaxy and the center of mass of its parent halo. We examine 23 relaxed galaxy clusters in a redshift range of 0.1 to 0.3 drawn from clusters in the Dark Energy Survey and the Sloan Digital Sky Survey which have archival Chandra X-ray data of sufficient depth for center and relaxation determination. We find that most clusters in our sample show non-zero offsets between the X-ray center, taken to be the centroid within the cluster core, and the central galaxy position.  All of the measured offsets are larger, typically by an order of magnitude, than the uncertainty in the X-ray position due to Poisson noise. In all but six clusters, the measured offsets are also larger than the estimated, combined astrometric uncertainties in the X-ray and optical positions.  A more conservative cut on concentration to select relaxed clusters marginally reduces but does not eliminate the observed offset.  With our more conservative sample, we find an estimated mean X-ray to central galaxy offset of $\mu = 5.5 \pm 1.0$ kpc. Comparing to recent simulations, this distribution of offsets is consistent with some level of dark matter self interaction, though further simulation work is needed to place constraints.
\end{abstract}




\section{Introduction}

The cold dark matter (CDM) paradigm postulates that dark matter is non-relativistic and collisionless. This model has been very successful in predicting the large-scale structure of the universe \citep{CDM, DMstructure}. 
However, potential discrepancies exist between theory and observations, particularly at smaller scales \citep[for reviews see, e.g.][]{noCDMrev,Bullock2017,BuckleyPeter}.  In addition, despite multi-pronged searches for CDM candidates, a conclusive non-gravitational signal has not been found, and basic tenants of the CDM paradigm, like dark matter's collisionless nature, are not yet strongly constrained \citep[e.g.][]{TulinYu}.

A generic possibility is that there may be undiscovered forces between dark matter particles; in this case, dark matter would possess non-zero self interactions and not be collisionless. Termed self-interacting dark matter (SIDM), this model was initially proposed as a solution to the core-cusp problem \citep{SIDMdiscovery}.  However, some form of self interaction is a generic beyond CDM possibility and a common feature of dark sector theories, and SIDM models predict potentially observable consequences for the shapes, densities, and substructure of dark matter halos \citep[see e.g. reviews by][]{TulinYu,Adhikari22}.

One important discrepancy between CDM simulations and observations is that the central circular velocity in galaxies is much lower than the velocities predicted by dark-matter only simulations \citep[e.g.][]{velDist1, velDist2}. Termed the "core/cusp problem", CDM predicts a sharper density peak in the centers of dark matter halos, scaling as $\rho_{DM} \propto r^{-1}$ \citep[e.g.][]{noCDM2, nfw1,nfw2}, than actually occurs with the data favoring constant density cores, particularly for low surface brightness and dwarf galaxies. \citep[e.g.][]{flores94,moore94,burkert95,blok01a,blok01b,kuzio08}. 
The inclusion of baryonic physics in hydrodynamical simulations and in particular supernova feedback goes a long way toward resolving the core/cusp discrepancy with simulations able to produce cores at least in some mass ranges \citep[e.g.][]{Governato10,Oh11,zolotov12}. However, it is unclear if baryonic effects alone can produce the full diversity of density profiles observed \citep[e.g.][]{Kuzio10,oman15}. See also \citet{TulinYu,Adhikari22} and references therein for discussion of the core/cusp problem and potential solutions.

Another possibility is that properties of the dark matter which deviate from the assumptions of CDM lead to the lower observed central densities.  As already mentioned, SIDM is one potential solution \citep[e.g.][]{SIDMdiscovery} in which self interactions lead to heat transfer from
outer, hotter regions to cooler, inner regions giving a uniform inner velocity dispersion and reduced central density.  Simulations have shown that SIDM plus baryons can naturally lead to the observed diversity in galaxy rotation curves and central densities \citep{Creasey17,kamada17,ren19,zavala19}. We note that other dark matter models have also been proposed that may produce cores and fix the observed small-scale issues, including warm dark matter \citep[e.g.][]{wdm2,viel13}, wave or fuzzy dark matter \citep[e.g.][]{fuzzy1, fuzzy2, hui21}, and superfluid dark matter \citep[e.g.][]{superfluid}.  

These different dark matter models may be distinguished by the scale dependence of their effects \citep{BuckleyPeter}.  In addition, on the particle theory side, many models for SIDM naturally lead to a dark matter self-interaction cross section over mass ($\sigma_\mathrm{DM}/m$) that depends on the relative velocity of the dark matter particles \citep[e.g][and references therein]{TulinYu,Adhikari22}, and a velocity-dependent cross section is preferred when reconciling dwarf galaxy observations with constraints on the scale of clusters of galaxies in an SIDM context \citep[e.g.][]{kaplinghat16}.  In this paper, we turn our attention to the most massive end of structure formation, clusters, and a recently proposed observational consequence of SIDM and central cores in general.



Using dark matter only simulations, \citet{thePaper} find that in SIDM models central cluster galaxies oscillate on long-lived orbits out to radii of 100 kpc or more and for up to Gyrs following a merger, likely even when clusters appear otherwise relaxed. In the CDM paradigm, the Central Cluster Galaxy (CCG) in a galaxy cluster will be constrained near the center by the high dark matter density with a position closely correlated to the center of mass of the dark matter halo. Conversely, in the SIDM paradigm the CCG will be statistically offset from the dark matter center of mass. The average offset from the center of mass is dependent on the halo core size, and thus for different cross-sections, the offset distribution will be different \citep{thePaper}. 
Including the effects of baryons using the BAHAMAS simulations, \citet{theOtherPaper} find much smaller, but observable offsets of the central cluster galaxies even in relaxed clusters with a median offset that increases with the SIDM cross section.  Using a sample of 10 strong lensing clusters, they derive a limit on the cross section of $\sigma/m < 0.4$ cm$^2$/g.



In this paper, we look for offsets of the central galaxy in 23 galaxy clusters selected to be both relaxed and X-ray bright. We estimate the position of the center of mass using the X-ray data, specifically the X-ray centroid within the core region, comparing them to the position of the CCG.  In Section 2, we discuss our methodology including cluster selection and relaxation criterion and cluster center determination. Section 3 presents our results, and in Section 4, we discuss the consequences and limitations of our analysis.
In this work, we assume a flat $\Lambda$CDM cosmology with $H_0 = 67.7$ km / (Mpc s) and $\Omega_M = 0.31$.

\section{Data Reduction and Methodology}

\subsection{Cluster Data}

We select clusters from two large area surveys, the Dark Energy Survey (DES) \citep{DES2005} and the Sloan Digital Sky Survey (SDSS) \citep{SDSS}, specifically DES Y3 Gold \citep{Y3Gold} and SDSS DR8 \citep{SDSSDR8}. 
Clusters were identified in the photometric survey data using the \textbf{red}-sequence \textbf{ma}tched-filter \textbf{P}robabilistic \textbf{Per}colation cluster finder (redMaPPer) algorithm \cite{redmapperI,redmapperSV}. RedMaPPer selects clusters based on overdensities of galaxies in color space.  More specifically, it identifies cluster members based on the red sequence, iteratively determining both the cluster center and the cluster red sequence.  Potential member galaxies in a given cluster are given a membership probability weighted by a matched filter based on color, magnitude, and separation distance from the estimated cluster center (taken to be the most probable identified central cluster galaxy). In this work, we use the SDSS redMaPPer v.6.3.1 and DES Y3 redMaPPer 6.4.22+2 catalogs with richness $\lambda>20$, and central galaxy positions were taken from these catalogs. Where both surveys overlap and had center galaxy positions for the same cluster, we removed the positions from the SDSS catalog and kept the positions from DES Year 3.


As described below, we use the X-ray brightness distribution to both select relaxed clusters and to estimate the cluster center of mass.  For this reason, our sample was limited to clusters with existing high spatial resolution Chandra X-ray data.  The Chandra data were reduced using the \textbf{M}ass \textbf{A}nalysis \textbf{T}ool for \textbf{Cha}ndra pipeline (MATCha) \citep{matcha}. 
For an input cluster catalog, MATCha reduces any existing Chandra data and determines cluster temperature and luminosity within several radii as well as finding the X-ray centroid and peak positions.  The SDSS X-ray analysis is described in \citet{matcha} and the DES Y3 analysis in \citet{Kelly22}.  In this work, we re-derive center locations starting from the reduced data as described in the next section.  Central galaxy positions are taken from the DES or SDSS catalogs.  In some cases, redMaPPer identifies the wrong galaxy as the central galaxy \citep{matcha,Zhang19}; for these clusters we identify the correct central galaxy based on proximity to the X-ray center and take the position of the correct central from the optical catalogs.

\subsection{Cluster Selection}

Beginning with the full DES and SDSS redMaPPer catalogs, we make a series of cuts to select a sample of clusters to perform our analysis on.  We need robustly X-ray detected clusters with high resolution, so we select clusters with redshifts between 0.1 and 0.3. The lower redshift limit is set by the requirement that the cluster X-ray emission fit within the Chandra field of view, while the upper redshift limit is set by the need to resolve position to within a few kpcs. For the signal-to-noise ratio of the X-ray data, we require a minimum ratio of 25 from the Chandra data in a 500 kpc radius region to ensure that the noise does not dominate the uncertainty in the center determination.  





As we will be using the X-ray emission as a tracer of the cluster center, we also select relaxed clusters where we expect the X-ray peak to trace fairly well the center of the gravitational potential. The selection of relaxed clusters is made based on the X-ray concentration.
The concentration is calculated by taking a ratio of the exposure-corrected photon counts within a radius of 15\% of the $R_{500}$ radius around the peak of X-ray emissions and in an annular region that extended from $0.15 R_{500}$ to $R_{500}$. We excluded observations in which the $R_{500}$ distance was greater than the area covered by the observation. 
The $R_{500}$ radius is the radius within which the average density of the cluster is 500 times larger than the critical density of the universe, and here we estimate $R_{500}$ from the X-ray temperature as in \cite{matcha}. We select clusters with concentrations of 0.5 or higher. This resulted in a final sample of 23 clusters.


Our definition of concentration differs somewhat from that of previous works as our definition relies on the $R_{500}$ radius. In particular, the criterion used by \cite{theOtherPaper} to select relaxed clusters in SIDM simulations was defined by a core radius of 100 kpc and an outer radius for the annulus of 300 kpc. As we are looking at galaxy clusters with a range of masses and size, defining the concentration based on an overdensity radius like $R_{500}$ allows us to more consistently compare statistics for both large and small galaxy clusters. 
Despite the difference in these criteria, the two concentration definitions are highly correlated; we find that the minimum concentration of 0.2 used in \cite{theOtherPaper} corresponds roughly to a concentration of 0.44 for our $R_{500}$ based concentration definition. After visual examination, we chose to make a slightly more conservative cut on concentration of 0.5 or greater. This gives a sample of 23 clusters; for comparison using the concentration definition of \citet{theOtherPaper} and their cut on concentration greater than 0.2 would have resulted in a sample of 28 clusters. Our visual examination showed that some clearly merging clusters would remain in the sample under the \citet{theOtherPaper} cut.  Even for our slightly more conservative cut, one clear merger remains, and in Section 3 we explore a cut on concentration greater than 0.6 to address this.

\subsection{Determination of Cluster Center}


The final part of the data reduction is the determination of the cluster center to be compared to the DES or SDSS central galaxy positions.  In this work, we define the cluster center to be the X-ray centroid determined within the cluster core, as described below.  As we have selected relaxed clusters, we expect the X-ray distribution to trace the underlying gravitational potential and thus mass distribution.


Specifically, we define the cluster center to be the centroid of the X-ray emission within a radius of 0.15$R_{500}$. This process requires that the initial center of the $R_{500}$ aperture be a good first-order approximation to the actual centroid. The MATCha algorithm approximates the X-ray peak position as the brightest pixel in the point-source subtracted, smoothed X-ray image. 
In order to improve this approximation, we created an iterative algorithm that measures the centroid of the 0.15$R_{500}$ aperture; after centroid determination we then use this centroid as the new center of the circle and recalculate the centroid until the difference between the iterations is less than two pixels.

In order to quantify the uncertainties in the centroid measurement, we added noise to the image and then remeasured the centroid and its offset from the central galaxy in the simulated image. To add noise to the image, we generated a random number on a Poisson distribution, where the mean value of the Poisson distribution for each pixel was taken to be the original pixel value in the image. We repeated this process 100 times, and took the median centroid to CCG offset as our accepted value, with a confidence interval defined as the range between the 16\textsuperscript{th} and 84\textsuperscript{th} smallest offsets.

\section{Results}


Results for the 23 clusters with concentrations greater than 0.5 are shown in Table~\ref{table:1}. The central galaxy offsets range from 1.5 kpc to 120 kpc, with most falling in between 4 and 15 kpc. The errors on these measurements are around an order of magnitude smaller than the measurements themselves, ranging from 0.1 to 1.0 kpc, with most falling between 0.1 and 0.4 kpc. Note that the size of the measured offsets are generally larger than but comparable to the pixel-size  and on-axis Chandra resolution of $\sim 0.5$ arcsec; we discuss the positional uncertainties further in Section 4. 

\begin{figure}
    \centering
    \includegraphics[width=0.9\linewidth]{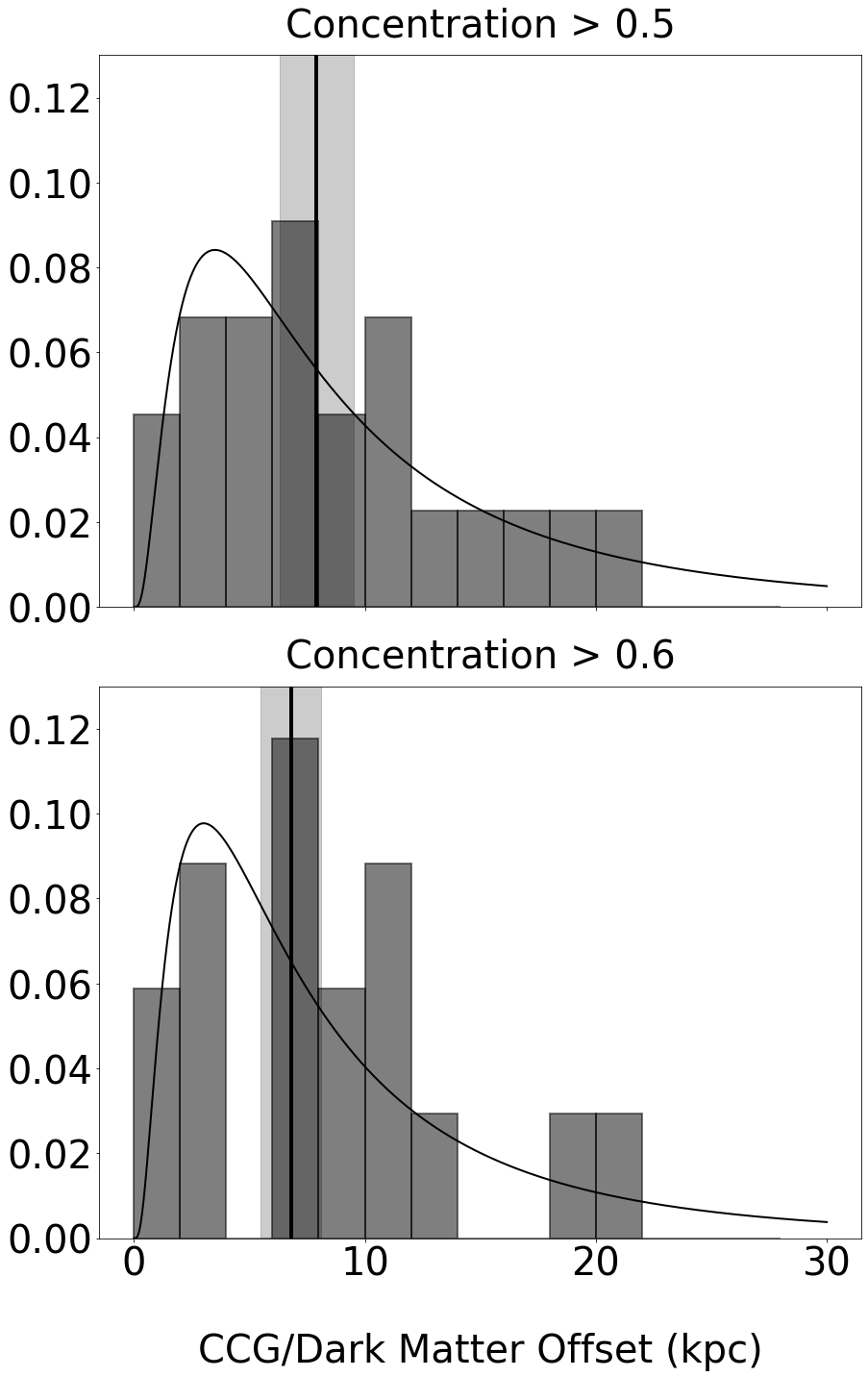}
    \caption{Histogram of CCG offsets from the calculated X-ray centroids for a concentration cut of 0.5 (top) and 0.6 (bottom). Also shown are the best fit log-normal distribution (black solid line) with $\mu = 7.9 \pm 1.6$ kpc and $\mu = 6.8 \pm 1.3$ kpc for the 0.5 and 0.6 concentration cuts, respectively. Vertical black lines and shaded region show the best-fit $\mu$ and 1$\sigma$ uncertainties estimated through bootstrapping.}
    \label{fig:hist}
\end{figure}

Following \citet{theOtherPaper}, we fit the central galaxy offset distribution to a log-normal probability density function with the form

$$ f(x) = \frac{1}{x \sigma \sqrt{2\pi}} \exp{\left(- \frac{1}{2\sigma^2} \ln^2{\left(\frac{x}{\mu}\right)}\right)}$$

using the Maximum Likelihood Estimator (MLE) from the Scipy library. Both the offset distribution and fit are shown in Figure~\ref{fig:hist}.
For the log-normal fit with our nominal concentration cut of 0.5, we find a mean value $\mu = 7.9 \pm 1.6$ kpc and $\sigma = 0.9 \pm 0.2$. Errors on the fit parameters were estimated using bootstrapping with a total of 100,000 trials. 
The parameter values, along, with their errors, are shown in Table~\ref{table:2}.

\begin{table*}
\begin{tabular}{cccccccccc}
\hline
Name & Catalog & Chandra & z & Offset & Error & $R_{500}$ & $\lambda$ & Concentration & Position Uncertainty\\
& & ObsID & & (kpc) & (kpc) & (Mpc) & & & (kpc)\\
\hline
RXCJ0232.2-4420   & DES Y3 & 4993  & 0.28 & 17.7  & 1.3 & 1.44 & 117.47 & 0.51 & 5.38 \\
MS 0906.5+1110    & SDSS   & 924   & 0.18 & 16.0  & 0.4 & 1.23 & 174.7  & 0.53 & 3.24 \\
A2445             & SDSS   & 12249 & 0.17 & 5.4   & 0.7 & 1.04 & 49.55  & 0.54 & 2.94 \\
RXC J0532.9-3701  & DES Y3 & 15112 & 0.15 & 4.7   & 0.7 & 1.51 & 199.43 & 0.55 & 2.72 \\
Abell S0592       & DES Y3 & 16572 & 0.25 & 119.6 & 0.4 & 1.51 & 96.37  & 0.57 & 4.56 \\
A853              & SDSS   & 12250 & 0.27 & 4.5   & 0.7 & 0.94 & 50.21  & 0.59 & 5.14 \\
RXJ1000.5+4409    & SDSS   & 9421  & 0.17 & 11.3  & 0.6 & 0.85 & 27.42  & 0.6  & 3.04 \\
RXCJ0220.9-3829   & DES Y3 & 9411  & 0.23 & 7.9   & 0.8 & 0.93 & 53.37  & 0.62 & 4.34 \\
RXCJ0307.0-2840   & DES Y3 & 9414  & 0.18 & 19.3  & 0.8 & 1.33 & 102.58 & 0.62 & 3.14 \\
A586              & SDSS   & 530   & 0.25 & 20.6  & 0.6 & 1.3  & 120.09 & 0.74 & 4.71 \\
RXC J2129.6+0005  & DES Y3 & 9370  & 0.25 & 3.1   & 0.5 & 1.21 & 76.58  & 0.8  & 4.73 \\
ABELL 2009        & SDSS   & 10438 & 0.19 & 7.0   & 0.4 & 1.24 & 91.82  & 0.91 & 3.39 \\
RXCJ0331.1-2100   & DES Y3 & 10790 & 0.25 & 3.3   & 0.5 & 1.01 & 64.75  & 0.99 & 4.73 \\
ZwCl 3146         & SDSS   & 909   & 0.27 & 13.0  & 0.2 & 1.19 & 73.53  & 1.0  & 5.09 \\
4C+55.16          & SDSS   & 4940  & 0.16 & 6.8   & 0.3 & 1.1  & 46.82  & 1.0  & 2.83 \\
Abell 1835        & SDSS   & 6880  & 0.18 & 11.4  & 0.1 & 1.35 & 134.28 & 1.07 & 3.17 \\
A383              & SDSS   & 2320  & 0.16 & 2.2   & 0.3 & 1.02 & 80.28  & 1.17 & 2.83 \\
RXJ1720.1+2638    & SDSS   & 4361  & 0.17 & 9.6   & 0.2 & 1.25 & 63.72  & 1.19 & 3.07 \\
ZwCl 0348         & DES Y3 & 10465 & 0.19 & 11.8  & 0.2 & 0.83 & 49.28  & 1.28 & 3.34 \\
ZwCl 2089         & SDSS   & 10463 & 0.29 & 1.6   & 0.2 & 0.88 & 27.08  & 1.29 & 5.56 \\
MS1455.0+2232     & SDSS   & 4192  & 0.24 & 7.7   & 0.2 & 1.04 & 54.92  & 1.35 & 4.48 \\
RXC J0132.6-0804  & DES Y3 & 16149 & 0.26 & 9.8   & 0.7 & 0.7  & 27.74  & 1.56 & 4.81 \\
ABELL 1204        & SDSS   & 2205  & 0.28 & 1.7   & 0.2 & 0.92 & 39.82  & 1.88 & 5.35 \\
\hline
\end{tabular}
\caption{Offsets of the central cluster galaxy from the X-ray center and associated errors. Tabulated is the cluster name (column 1), the survey from which the cluster was drawn (column 2), the Chandra obsID (column 3), the redmaPPer redshift (column 4), the CCG to X-ray offset (column 5), the uncertainty in the offset due to Poisson noise (column 6), the X-ray concentration (column 7), and the physical size corresponding to the estimated, combined positional uncertainties of the X-ray and optical imaging (column 8). Note that the positional uncertainties for these clusters are all larger than the measurement uncertainties due to Poisson noise. The measured offsets are all less than roughly 20 kpc with the exception of Abell S0592 with an offset of 119.6 kpc. Upon further inspection of this outlier, it was determined to be a merging cluster.}
\label{table:1}
\end{table*}

\begin{table}
\begin{tabular}{ccccc}
\hline
Concentration & $\mu$ & Bootstrap & $\sigma$ & Bootstrap\\
Minimum & (kpc) & Error & & Error\\
\hline
0.5 & 7.9 & 1.6 & 0.9 & 0.2 \\
0.6 & 6.8 & 1.3 & 0.8 & 0.1 \\
\hline
\end{tabular}
\caption{Results from fitting the offsets to a log normal probability density function with parameters $\mu$ for the mean value and $\sigma$ for the variance. Errors on the model parameters were found via bootstrapping.}
\label{table:2}
\end{table}

Of the nominal sample of 23 clusters the measured offsets are typically 20 kpc or less with the exception of Abell S0592. This cluster has an offset that is one to two orders of magnitude larger than the rest of the sample. 
Abell S0592 is a known merger \citep{AS592}, and an examination the X-ray images of showed two X-ray substructures and two possible CCGs in the cluster. For a log-normal fit of the data with just this outlier removed, the mean value of this set of clusters is 7.0 kpc. However, the fact that this concentration let in a merger at all implies that the 0.5 concentration cut was not conservative enough to remove merging clusters in non-simulated data. 

Accordingly, we re-cut the data with a concentration of 0.6, which removed the merging cluster along with five other clusters. With this cut, the offsets range between 1.5 and 20.5 kpc, with most offsets falling between 4 and 15 kpc. A log-normal fit to the offset distribution for this more conservative sample resulted in a mean value of 6.8 $\pm$ 1.3 kpc; the distribution and fit are both shown in the lower panel of Figure~\ref{fig:hist}.

\section{Discussion}


Our results indicate a non-zero offset between the central galaxy and X-ray center in our relaxed cluster sample even for our more conservative cut on concentration.  As already noted, the offsets in all cases are significantly larger than the estimated uncertainty from Poisson noise.  We now consider the accuracy of the positions of both Chandra X-ray sources and the central galaxies.  For Chandra, the absolute positional accuracy when comparing measured point source X-ray centroids to optical or radio counterparts with well-measured positions is 
$\sim0.7"$ (68\%)\footnote{Chandra Proposers' Observatory Guide, Cycle 24:\\ https://cxc.harvard.edu/proposer/POG/html/index.html}.
In comparison, the DES Y3 Gold average absolute astrometric accuracy is 0.158" \citep{Y3Gold}.  Comparing the SExtractor \citep{sextractor} and ngmix \citep{Y3Gold} estimated positions for the DES central galaxies in our sample, we find an average difference of 0.18" and use this as an estimate of the modeling uncertainties.  Adding these in quadrature, gives an estimated central galaxy position uncertainty of 0.24".  Taken together the X-ray and optical positional uncertainties imply an uncertainty on the measured offsets of 0.74".
 
In all but six cases, the measured offsets are larger than the positional uncertainty at the cluster redshift, though for a few additional clusters the offset and resolution are comparable.  For the 17 clusters with concentration greater than 0.6, the average positional uncertainty is 4.0 kpc, ranging from $2.7-5.6$ kpc.  Subtracting the average uncertainty in quadrature from the measured $\mu  = 6.8 \pm 1.3$ kpc, gives $5.5^{+1.5}_{-1.7}$ kpc. For the less conservative concentration cut of 0.5 after accounting for the positional uncertainty, we find an average offset of $\mu  = 6.8^{+1.8}_{-1.9}$ kpc.

In comparison, \citet{theOtherPaper} similarly fit a log-normal to the distribution of CCG offsets for relaxed clusters in their simulations; they find that SIDM with a self-interacting cross-section of $1.0$ cm$^2$/g has a $\mu$ value of $8.6 \pm 0.7$ kpc, while a self-interacting cross-section of 0.3 cm$^2$/g gives a $\mu$ value of $6.1 \pm 0.7$ kpc and CDM a $\mu$ value of $3.8 \pm 0.7$ kpc.  These mean offsets are much smaller than the offsets predicted by the simulations from \cite{thePaper}, which were dark matter only.  However, the quoted offsets from \citet{theOtherPaper} are the initial results from their simulations before they accounted for numerical effects. They attempt to model the effects of the limited resolution of their simulations (softening length $\epsilon = 4 h^{-1}$ kpc) using much smaller cluster samples re-simulated at higher resolution for two models, CDM and SIDM with cross-section $1.0$ cm$^2$/g.  Their nominal results give a resolution corrected $\mu = 0.8^{+0.9}_{-0.8}$ kpc for CDM and $\mu = 2.3^{+1.8}_{-0.7}$ kpc for SIDM $1.0$ cm$^2$/g.  They go on to derive a limit on the cross-section of $\sigma/m < 0.4$ cm$^2$/g using the CCG offsets in 10 strong lensing clusters \citep{theOtherPaper}.  However, these numbers and comparisons to our results come with several important caveats.  In particular, the resolution correction relies on small samples ($\sim20$) of clusters run at higher resolution for only two different dark matter models.  In addition, the high resolution simulations lead to different inner stellar density profiles in the simulated clusters mixing baryonic and dark matter effects.  The model for the resolution correction used is thus not well constrained nor particularly well motivated.  Finally, our results are for the offset between the X-ray center and central galaxy, which is not directly what is measured in the simulations.  In addition, the BAHAMAS simulations used in \citet{theOtherPaper} represent one implementation of baryonic physics, and more work is needed on the simulation side to understand how variations in the baryonic physics affect the central galaxy offsets.

In general, our measurement of central galaxy offsets on the scale of several kpc are consistent with the expectations for an non-negligible SIDM cross-section on the order of $\sim1.0$ cm$^2$/g and mildly in tension with the resolution-corrected CDM expectation \citep{theOtherPaper}.  However, given the caveats above we cannot rule out CDM. More work on the simulation side is needed to fully utilize and interpret our results, which is beyond the scope of the current paper.
Our results are also compatible with current constraints on the SIDM cross-section.  For example, \citet{Sagunski2021} find an upper limit on a velocity dependent cross section in clusters of $\sigma/m < 0.35$ cm$^2$/g considering the measured inner densities of strong lensing clusters and groups \citep[see also][]{Newman2013}. Considering a range of observations, \citet{TulinYu} conclude that a cross-section $\sigma/m \sim 0.5 - 1$ cm$^2$/g could resolve small scale structure problems while staying roughly consistent with large scale structure constraints. However, they find that dwarf and low surface brightness galaxy observations imply cross sections of $\sigma/m > 1$ cm$^2$/g, and generally some velocity dependence of the self-interaction cross section is preferred to reconcile cluster observations with smaller scales. 


\section{Conclusions}

We use a combination of Chandra X-ray data and optical imaging from the DES and SDSS surveys to investigate the presence of offsets between cluster central galaxies and the cluster gravitational center that are predicted to exist in models of dark matter including dark matter self interactions.  We use a cut on X-ray concentration to select relatively relaxed clusters and a redshift cut to ensure sufficient spatial resolution, resulting in a sample of 23 clusters. 

We measured the offset between the centroid of X-ray emission and the central cluster galaxy. Modelling the distribution of offsets with a log-normal distribution, we found the mean value of the offset to be $\mu = 7.9 \pm 1.6$ kpc.  As our initial concentration cut allowed one clear merger to remain in the sample, we also explored a more restrictive cut, resulting in a sample of 17 clusters. We found the mean value of the offset for this sample to be $\mu = 6.8 \pm 1.3$ kpc. Our results indicate non-zero offsets for most clusters in the sample.  Using Monte Carlo resimulations of the noise, we find that the uncertainty in the X-ray positions due to noise are typically an order of magnitude smaller than the measured offsets, with an average uncertainty of 0.5 kpc. Uncertainties in the absolute astrometry of both the X-ray and optical observations are larger, but still lower than the measured offsets for most of the clusters in our sample.  Taking the more conservative concentration cut and accounting for the average positional uncertainties in both the X-ray and galaxy positions, results in an estimated mean offset of $\mu = 5.5^{+1.5}_{-1.7}$ kpc.

Regardless of the concentration cut made, our results are consistent with some level of dark matter self interaction of $\sigma/m \sim 1.0$ cm$^2$/g when compared to the simulated results for relaxed clusters from \citet{theOtherPaper}, but we also cannot completely rule out CDM given uncertainties in the simulated results and our use of the X-ray center as a proxy for the dark matter position. 
In the future, these constraints can be improved by expanding the cluster sample size, in particular with additional high-resolution X-ray observations of DES and SDSS clusters, and through the development of simulations which more directly simulate the measurements and cuts made here.

\section*{Acknowledgements}
This work was supported by the U.S. Department of Energy under Award Number DE-SC0010107.  Funding for the DES Projects has been provided by the U.S. Department of Energy, the U.S. National Science Foundation, the Ministry of Science and Education of Spain, 
the Science and Technology Facilities Council of the United Kingdom, the Higher Education Funding Council for England, the National Center for Supercomputing 
Applications at the University of Illinois at Urbana-Champaign, the Kavli Institute of Cosmological Physics at the University of Chicago, 
the Center for Cosmology and Astro-Particle Physics at the Ohio State University,
the Mitchell Institute for Fundamental Physics and Astronomy at Texas A\&M University, Financiadora de Estudos e Projetos, 
Funda{\c c}{\~a}o Carlos Chagas Filho de Amparo {\`a} Pesquisa do Estado do Rio de Janeiro, Conselho Nacional de Desenvolvimento Cient{\'i}fico e Tecnol{\'o}gico and 
the Minist{\'e}rio da Ci{\^e}ncia, Tecnologia e Inova{\c c}{\~a}o, the Deutsche Forschungsgemeinschaft and the Collaborating Institutions in the Dark Energy Survey. 

The Collaborating Institutions are Argonne National Laboratory, the University of California at Santa Cruz, the University of Cambridge, Centro de Investigaciones Energ{\'e}ticas, 
Medioambientales y Tecnol{\'o}gicas-Madrid, the University of Chicago, University College London, the DES-Brazil Consortium, the University of Edinburgh, 
the Eidgen{\"o}ssische Technische Hochschule (ETH) Z{\"u}rich, 
Fermi National Accelerator Laboratory, the University of Illinois at Urbana-Champaign, the Institut de Ci{\`e}ncies de l'Espai (IEEC/CSIC), 
the Institut de F{\'i}sica d'Altes Energies, Lawrence Berkeley National Laboratory, the Ludwig-Maximilians Universit{\"a}t M{\"u}nchen and the associated Excellence Cluster Universe, 
the University of Michigan, NSF's NOIRLab, the University of Nottingham, The Ohio State University, the University of Pennsylvania, the University of Portsmouth, 
SLAC National Accelerator Laboratory, Stanford University, the University of Sussex, Texas A\&M University, and the OzDES Membership Consortium.

Based in part on observations at Cerro Tololo Inter-American Observatory at NSF's NOIRLab (NOIRLab Prop. ID 2012B-0001; PI: J. Frieman), which is managed by the Association of Universities for Research in Astronomy (AURA) under a cooperative agreement with the National Science Foundation.

The DES data management system is supported by the National Science Foundation under Grant Numbers AST-1138766 and AST-1536171.
The DES participants from Spanish institutions are partially supported by MICINN under grants ESP2017-89838, PGC2018-094773, PGC2018-102021, SEV-2016-0588, SEV-2016-0597, and MDM-2015-0509, some of which include ERDF funds from the European Union. IFAE is partially funded by the CERCA program of the Generalitat de Catalunya.
Research leading to these results has received funding from the European Research
Council under the European Union's Seventh Framework Program (FP7/2007-2013) including ERC grant agreements 240672, 291329, and 306478.
We  acknowledge support from the Brazilian Instituto Nacional de Ci\^encia
e Tecnologia (INCT) do e-Universo (CNPq grant 465376/2014-2).

This manuscript has been authored by Fermi Research Alliance, LLC under Contract No. DE-AC02-07CH11359 with the U.S. Department of Energy, Office of Science, Office of High Energy Physics.


\bibliographystyle{mnras}
\bibliography{paper}

\section{Affiliations}
$^{1}$ University of California, Santa Cruz, Santa Cruz, CA 95064, USA\\
$^{2}$ Santa Cruz Institute for Particle Physics, Santa Cruz, CA 95064, USA\\
$^{3}$ Institute of Space Sciences (ICE, CSIC),  Campus UAB, Carrer de Can Magrans, s/n,  08193 Barcelona, Spain\\
$^{4}$ San Francisco State University, 1600 Holloway Ave, San Francisco, CA 94132\\
$^{5}$ IISER Pune, Dr. Homi Bhabha Road,Pune 411008, India\\
$^{6}$ University Observatory, Faculty of Physics, Ludwig-Maximilians-Universitat, Scheinerstr. 1, 81679 Munich, Germany\\
$^{7}$ SLAC National Accelerator Laboratory, 2575 Sand Hill Road, Menlo Park, CA 94025\\
$^{8}$ Jet Propulsion Laboratory, California Institute of Technology, 4800 Oak Grove Dr., Pasadena, CA 91109, USA\\
$^{9}$ Department of Physics and Astronomy, Clemson University, Kinard Lab of Physics, Clemson, SC 29634-0978, US\\
$^{10}$ Laborat\'orio Interinstitucional de e-Astronomia - LIneA, Rua Gal. Jos\'e Cristino 77, Rio de Janeiro, RJ - 20921-400, Brazil\\
$^{11}$ Department of Physics, University of Michigan, Ann Arbor, MI 48109, USA\\
$^{12}$ Institute of Cosmology and Gravitation, University of Portsmouth, Portsmouth, PO1 3FX, UK\\
$^{13}$ CNRS, UMR 7095, Institut d'Astrophysique de Paris, F-75014, Paris, France\\
$^{14}$ Sorbonne Universit\'es, UPMC Univ Paris 06, UMR 7095, Institut d'Astrophysique de Paris, F-75014, Paris, France\\
$^{15}$ Department of Physics \& Astronomy, University College London, Gower Street, London, WC1E 6BT, UK\\
$^{16}$ Kavli Institute for Particle Astrophysics \& Cosmology, P. O. Box 2450, Stanford University, Stanford, CA 94305, USA\\
$^{17}$ SLAC National Accelerator Laboratory, Menlo Park, CA 94025, USA\\
$^{18}$ Instituto de Astrofisica de Canarias, E-38205 La Laguna, Tenerife, Spain\\
$^{19}$ Universidad de La Laguna, Dpto. Astrofísica, E-38206 La Laguna, Tenerife, Spain\\
$^{20}$ Center for Astrophysical Surveys, National Center for Supercomputing Applications, 1205 West Clark St., Urbana, IL 61801, USA\\
$^{21}$ Department of Astronomy, University of Illinois at Urbana-Champaign, 1002 W. Green Street, Urbana, IL 61801, USA\\
$^{22}$ Physics Department, William Jewell College, Liberty, MO, 64068\\
$^{23}$ Astronomy Unit, Department of Physics, University of Trieste, via Tiepolo 11, I-34131 Trieste, Italy\\
$^{24}$ INAF-Osservatorio Astronomico di Trieste, via G. B. Tiepolo 11, I-34143 Trieste, Italy\\
$^{25}$ Institute for Fundamental Physics of the Universe, Via Beirut 2, 34014 Trieste, Italy\\
$^{26}$ Hamburger Sternwarte, Universit\"{a}t Hamburg, Gojenbergsweg 112, 21029 Hamburg, Germany\\
$^{27}$ School of Mathematics and Physics, University of Queensland,  Brisbane, QLD 4072, Australia\\
$^{28}$ Department of Physics, IIT Hyderabad, Kandi, Telangana 502285, India\\
$^{29}$ Institute of Theoretical Astrophysics, University of Oslo. P.O. Box 1029 Blindern, NO-0315 Oslo, Norway\\
$^{30}$ Fermi National Accelerator Laboratory, P. O. Box 500, Batavia, IL 60510, USA\\
$^{31}$ Kavli Institute for Cosmological Physics, University of Chicago, Chicago, IL 60637, USA\\
$^{32}$ Instituto de Fisica Teorica UAM/CSIC, Universidad Autonoma de Madrid, 28049 Madrid, Spain\\
$^{33}$ Institut de F\'{\i}sica d'Altes Energies (IFAE), The Barcelona Institute of Science and Technology, Campus UAB, 08193 Bellaterra (Barcelona) Spain\\
$^{34}$ University Observatory, Faculty of Physics, Ludwig-Maximilians-Universit\"at, Scheinerstr. 1, 81679 Munich, Germany\\
$^{35}$ Center for Cosmology and Astro-Particle Physics, The Ohio State University, Columbus, OH 43210, USA\\
$^{36}$ Department of Physics, The Ohio State University, Columbus, OH 43210, USA\\
$^{37}$ Center for Astrophysics $\vert$ Harvard \& Smithsonian, 60 Garden Street, Cambridge, MA 02138, USA\\
$^{38}$ Australian Astronomical Optics, Macquarie University, North Ryde, NSW 2113, Australia\\
$^{39}$ Lowell Observatory, 1400 Mars Hill Rd, Flagstaff, AZ 86001, USA\\
$^{40}$ George P. and Cynthia Woods Mitchell Institute for Fundamental Physics and Astronomy, and Department of Physics and Astronomy, Texas A\&M University, College Station, TX 77843,  USA\\
$^{41}$ Centro de Investigaciones Energ\'eticas, Medioambientales y Tecnol\'ogicas (CIEMAT), Madrid, Spain\\
$^{42}$ Instituci\'o Catalana de Recerca i Estudis Avan\c{c}ats, E-08010 Barcelona, Spain\\
$^{43}$ Observat\'orio Nacional, Rua Gal. Jos\'e Cristino 77, Rio de Janeiro, RJ - 20921-400, Brazil\\
$^{44}$ Department of Astrophysical Sciences, Princeton University, Peyton Hall, Princeton, NJ 08544, USA\\
$^{45}$ Department of Physics, University of Genova and INFN, Via Dodecaneso 33, 16146, Genova, Italy\\
$^{46}$ Department of Physics and Astronomy, Pevensey Building, University of Sussex, Brighton, BN1 9QH, UK\\
$^{47}$ School of Physics and Astronomy, University of Southampton,  Southampton, SO17 1BJ, UK\\
$^{48}$ Instituto de F\'isica Gleb Wataghin, Universidade Estadual de Campinas, 13083-859, Campinas, SP, Brazil\\
$^{49}$ Computer Science and Mathematics Division, Oak Ridge National Laboratory, Oak Ridge, TN 37831\\
$^{50}$ National Center for Supercomputing Applications, 1205 West Clark St,Urbana, IL 61801, USA\\
$^{51}$ Lawrence Berkeley National Laboratory, 1 Cyclotron Road, Berkeley, CA 94720, USA\\
$^{52}$ Max Planck Institute for Extraterrestrial Physics, Giessenbachstrasse, 85748 Garching, Germany\\
$^{53}$ Universit\"ats-Sternwarte, Fakult\"at f\"ur Physik, Ludwig-Maximilians Universit\"at M\"unchen, Scheinerstr. 1, 81679 M\"unchen, Germany\\

\bsp	
\label{lastpage}
\end{document}